\documentclass[twocolumn,aps,showpacs,amsmath,superscriptaddress]{revtex4-2}
\usepackage[latin9]{luainputenc}
\usepackage{mathtools}
\usepackage{amsmath}
\usepackage{amssymb}
\usepackage{graphicx}
\usepackage[symbol]{footmisc}
\usepackage[pagewise]{lineno}

\makeatletter

\usepackage{dcolumn}
\usepackage{bm}
\usepackage{ifpdf}
\usepackage{lineno}\usepackage{amsfonts}
\usepackage{bbm}
\usepackage{mathrsfs}
\usepackage{upgreek}
\usepackage{epstopdf}
\usepackage{setspace}
\usepackage[matrix,frame,arrow]{xypic}
\usepackage{rotating}
\usepackage{float}
\usepackage{hyperref}
\usepackage{color}
\definecolor{med-blue}{RGB}{25,25,112}
\hypersetup{colorlinks, linkcolor={blue},citecolor={blue}, urlcolor={blue}}

\newcommand{\ket}[1]{\vert{#1}\rangle}

\newcommand{\expec}[1]{\langle{#1}\rangle}

\makeatother

\begin{document}
	
    \title{Spin bath mediated long-lived coherent oscillations of NV centers in diamond}
	\author{Akshat Rana}
	\affiliation{Department of Physics, Bennett University, Greater Noida 201310, India}
	\author{Pooja Lamba}
	\affiliation{Department of Physics, Bennett University, Greater Noida 201310, India}
	\author{Basanta Mistri}
	\affiliation{Department of Physics, IIT Madras, Chennai 600036, India}
	\author{Dieter Suter}
	\affiliation{Fakult{\"a}t Physik, Technische Universit{\"a}t Dortmund, D-44221 Dortmund, Germany}
	\author{Siddharth Dhomkar}
	\email{sdhomkar@physics.iitm.ac.in}
	\affiliation{Department of Physics, IIT Madras, Chennai 600036, India}
	\affiliation{Center For Quantum Information, Communication And Computing, IIT Madras, Chennai 600036, India}
	\affiliation{London center for Nanotechnology, UCL, London WC1H0AH, UK}
	\author{Rama K. Kamineni}
	\email{koti.kamineni@gmail.com}
	\affiliation{Department of Physics, Bennett University, Greater Noida 201310, India}
	\affiliation{SIAS, Krea University, Sri City 517646, India}

	\date{\today}
	\begin{abstract}
		{Decoherence is the biggest bottleneck in all quantum technologies. For nitrogen-vacancy (NV) centers in diamond, the loss of coherence is caused by the electron and nuclear spin bath of the diamond lattice. Here, we demonstrate that the spin bath - that typically causes decoherence - entangles the spin states of the NV electron and the host $^{14}$N nucleus. The many-body interaction between the $^{14}$N nucleus - electron - bath spins at an energy level anti-crossing occurring for an applied magnetic field orientation perpendicular to the NV axis is responsible for this effect. This is observed experimentally in NV ensembles via electron spin-echo measurements, where the echo envelope is modulated at the frequency of a $^{14}$N nuclear spin transition. Using numerical simulations, we show that the spin bath coupling to the NV centers is essential for observing this modulation. Due to the zero first-order Zeeman effect at the anti-crossing, the observed oscillations have long spin-echo coherence times, 2--3 times those at the parallel magnetic field orientation. The oscillation frequency is highly stable and robust against environmental fluctuations. These findings provide new opportunities for fundamental studies of many-body physics and quantum sensing.
			
		} 
	\end{abstract}
	
	
	\maketitle

Long coherence times are one of the fundamental prerequisites of quantum technologies \cite{Zurek1991Decoherence,Dowling2003SecQRevol}. For fault-tolerant quantum computing, coherence times of the qubits must be much longer than the gate operation times \cite{Loss1998QuantumDots,DiVincenzo2000Criteria}. Quantum networks require high-fidelity and long-lived memories for storing and synchronizing quantum information \cite{Briegel1998QRepeaters,Lvovsky2009OpticalMemory}. In quantum sensing, the sensitivity of the probing qubit depends on its ability to maintain superposition states for a long duration \cite{DegenRevSensing2017, Barry_SensRev2020}. 
In solid-state spin systems, loss of coherence is caused by magnetic field fluctuations arising from the surrounding electron and nuclear spin bath. Conventionally,  the eigenstates are taken as those of the pure system Hamiltonian, while the system-environment interaction only modifies the dynamics through time-dependent perturbations. In this paper, we focus on a scenario in which the spin bath not only causes decoherence but also modifies the system's eigenstates.

Optical spin initialization and readout, long coherence times, and miniature sensor size make NV centers in diamond useful for several quantum technologies including quantum information processing and metrology \cite{RevDegen,Rovny2024DiaSensManyBody,Katsumi2025HybDPhotonics}. 
Decoherence of NV centers is caused by the electron spins of paramagnetic impurities and the $^{13}$C nuclear spins of the diamond lattice \cite{Park2022NVDecoh}. In ultra-pure diamond crystals containing low concentration of paramagnetic impurities, the spin bath is dominated by $^{13}$C nuclear spins \cite{Zhao2012Decoh}.
Dynamical decoupling can extend the coherence time of NV centers \cite{deLange2010UD,Suter2016Colloquium}, which depends on the angle between the NV axis and the applied magnetic field.
It has been reported that the coherence time of NV centers in ultra-pure diamonds is maximum when the field is aligned with the NV axis \cite{MazeDecoh,WalsworthDecoh}. In diamond containing a very high concentration of paramagnetic impurities, the coherence time improves when the field is oriented perpendicular to the NV axis \cite{BajajDecoh}. However, in this crystal, the coherence time is only a few microseconds, limiting its applications. In this diamond, spin-echo modulations associated with the $^{14}$N nuclear quadrupolar interaction have also been observed under a transverse magnetic field \cite{Bajaj14Nquadrapole}.

The simplest dynamical-decoupling pulse sequence is the spin-echo, which employs a $\pi$-pulse at the center of the free-evolution period \cite{Hahn1950SpinEcho}. When the spin bath is dominated by $^{13}$C nuclear spins, the electron spin-echo signal exhibits characteristic decays and revivals at half the $^{13}$C Larmor precession frequency \cite{Chil2006Sci}.  
The spin-1 character of the NV electron spin and the anisotropic nature of the hyperfine interaction with the spin-1/2 $^{13}$C nuclei are crucial for understanding the echo revivals. The quantization axis of the bath nuclear spins depends on the state of the NV electron spin, conditioning the nuclear spin states on the electron spin state.

The host $^{14}$N atom of the NV center possesses a spin-1 nucleus with a strong nuclear quadrupolar coupling, whose principal axis coincides with the NV axis. Since the $^{14}$N atom is positioned on the NV axis, its hyperfine tensor is axially symmetric. The Zeeman interaction of the $^{14}$N nuclear spin is typically small compared to the quadrupolar coupling. When a static magnetic field is applied parallel to the NV axis, the quantization axes of both electron and $^{14}$N nuclear spins align with the NV axis. Consequently, the nuclear spin eigenbasis is common to all the electron spin manifolds. Here, we observe that the $^{14}$N nuclear spin state is conditioned on the NV electron spin state when a magnetic field in the range 6--21 mT is applied perpendicular to the NV axis. Interestingly, this occurs due to the hyperfine coupling between the electron spin and the bath $^{13}$C nuclear spins, in particular, due to the off-diagonal elements of the hyperfine tensor. 
As a characteristic of this $^{14}$N nucleus - NV electron - $^{13}$C spin bath many-body interaction, the electron spin-echo signal is modulated by a $^{14}$N nuclear spin transition, whose frequency depends on the quadrupolar interaction, the transverse $^{14}$N hyperfine parameter, and the magnetic field. These electron spin-echo modulations also have long coherence times, 2--3 times the $T_2$ values at the parallel field, due to the zero first-order Zeeman (ZEFOZ) shift occurring at the energy level anti-crossings induced by the transverse field. 

\begin{figure}
	\centering \includegraphics[width=8.5cm]{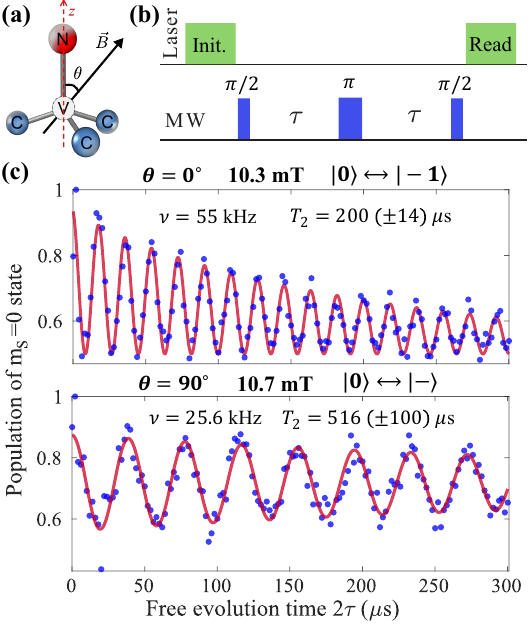} 
	\caption{(a) Schematic of the NV center and static magnetic field. (b) Electron spin-echo pulse sequence. (c) Measured spin-echo signals for the transitions $\ket{0}\leftrightarrow\ket{-1}$ and $\ket{0}\leftrightarrow\ket{-}$. The former is fitted to $a\cos^2(\pi\nu\,2\tau)\exp[-(2\tau/T_2)^n]+b$ with $n\approx 1$, and the latter to $a\cos(2\pi\nu\,2\tau)\exp(-2\tau/T_2)+b$.}
	\label{fig1} 
\end{figure}


The ground-state Hamiltonian of the NV center in an applied magnetic field is 
${\cal H}_1=DS_z^2 - \gamma_e \vec{B}\cdot\vec{S}$,
where $\vec{B}=B(\sin \theta \cos \phi, \sin \theta \sin \phi, \cos \theta)$ and $\vec{S}=(S_x, S_y, S_z)$ are the magnetic field and spin-1 angular momentum vectors respectively. We set $h=1$ and use frequency units for energy. $D=2870$ MHz and $\gamma_e=-28025$ MHz/T are the zero-field splitting and gyromagnetic ratio of the electron spin respectively.
When the magnetic field is applied perpendicular to the NV axis, the energy level anti-crossing \cite{PoojaVecSens2024, Akshat2025MWsens} strongly mixes the $\ket{-1}$ and $\ket{1}$ states, leading to the approximate eigenstates $\ket{0}$, $\ket{-}$, and $\ket{+}$, where $\ket{\pm}=\frac{\ket{-1}\pm \ket{1}}{\sqrt{2}}$, with $0$, $-1$, $1$ denoting the eigenvalues of $S_z$.

We perform electron spin-echo measurements on two diamond crystals containing NV center ensembles, where the spin bath is dominated by $^{13}$C nuclear spins.
Fig. \ref{fig1}(b) shows the pulse sequence used to drive the transitions $\ket{0}\leftrightarrow\ket{-1}$ at $\theta=0^\circ$ and $\ket{0}\leftrightarrow\ket{-}$ at $\theta=90^\circ$. The corresponding results, obtained at magnetic field strengths of 10.3 and 10.7 mT respectively, are shown in Fig. \ref{fig1}(c). For the parallel field, the signal shows characteristic decays and revivals at one-half of the $^{13}$C Larmor precession frequency ($\frac{1}{2} \gamma_C B$, where $\gamma_C$ is the gyromagnetic ratio of the $^{13}$C nuclear spin). The $T_2$ coherence time is measured to be 200 ($\pm 14$) $\mu$s. For the transverse field and the $\ket{0}\leftrightarrow\ket{-}$ transition, the signal shows oscillations at a frequency of 25.6 kHz, which is quite different from the $\frac{1}{2} \gamma_C B=57$ kHz. The $T_2$ value for these oscillations is 516 ($\pm 100$) $\mu$s. Similar oscillations are observed for the $\ket{0}\leftrightarrow\ket{+}$ transition as well. 
However, the electron spin transition $\ket{-}\leftrightarrow\ket{+}$, which is allowed for transverse field \cite{PoojaVecSens2024}, exhibit modulation at $\frac{1}{2} \gamma_C B$ (Supplementary Section S3). The presence of 25.6 kHz electron spin-echo envelope modulations (ESEEM) at 10.7 mT for the transitions $\ket{0}\leftrightarrow\ket{-}$ and $\ket{0}\leftrightarrow\ket{+}$, and their absence for the transition $\ket{-}\leftrightarrow\ket{+}$ indicates that the transition responsible for these modulations must be in the $\ket{0}$ subspace. 

\begin{figure}
	\centering \includegraphics[width=8.8cm]{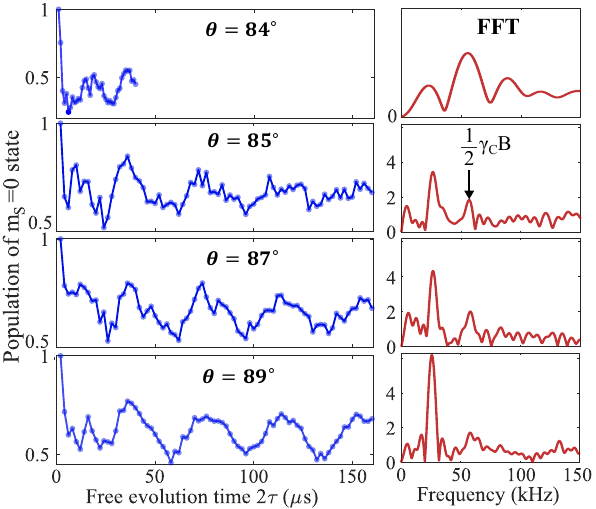} 
	\caption{Electron spin-echo signals and their Fourier transforms (FFTs) of the $\ket{0}\leftrightarrow\ket{-}$ transition for magnetic field orientations orthogonal to the NV-axis. Here, $B=10.7$ mT.}
	\label{fig:EchoVsTh} 
\end{figure}

\begin{figure}
	\centering \includegraphics[width=8.5cm]{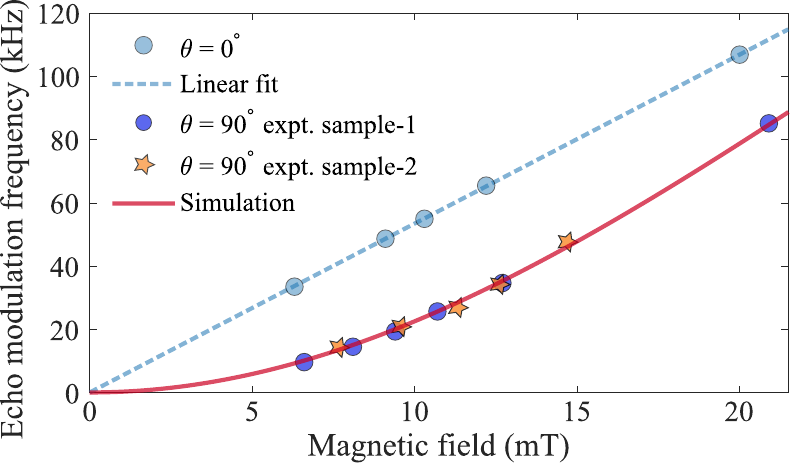} 
	\caption{Frequencies of spin-echo modulations versus magnetic field strength for parallel ($\theta=0^{\circ}$) and transverse ($\theta=90^{\circ}$) fields. $\theta=0^\circ$ data follow $\frac{1}{2}\gamma_C B$, whereas $\theta=90^\circ$ data are fitted to a $^{14}$N transition frequency ($\ket{e^-,{^{14}N},{^{13}C}}=\ket{0,+,\alpha_1}\leftrightarrow\ket{0,-,\alpha_1}$, where $\ket{\pm}=\frac{\ket{-1}\pm \ket{1}}{\sqrt{2}}$) and $\ket{\alpha_1}$ is a superposition of $\ket{-\frac{1}{2}}$ and $\ket{\frac{1}{2}}$). Error bars for the modulation frequencies are smaller than the data-point markers.}
	\label{fig:FreqVsField} 
\end{figure}


To understand the ESEEM oscillations further, we measure spin-echo signals for a range of $\theta$ values near the transverse field orientation. The resulting signals and their Fourier transforms are shown in Fig. \ref{fig:EchoVsTh}. At $\theta=84^\circ$, the signal decays very quickly and it shows very weak modulation associated with $^{13}$C Larmor precession. As $\theta$ increases to $85^\circ$, the 25.6 kHz ESEEM oscillations start to appear. 
The amplitude of these oscillations and their coherence time increase as $\theta$ approaches $90^\circ$, while the modulation associated with $^{13}$C Larmor precession becomes negligible. This indicates that both enhanced $T_2$ and ESEEM oscillations originate due to the transverse magnetic field. 

The echo modulation frequency as a function of the magnetic field strength for both parallel and transverse orientations are shown in Fig. \ref{fig:FreqVsField}. As expected, the echo revival frequencies for $\theta=0^\circ$ can be fitted to $\frac{1}{2} \gamma_C B$. However, the ESEEM  frequency at $\theta=90^\circ$ increases non-linearly with the field strength.


When a magnetic field is applied perpendicular to the NV-axis, the expectation values of the electron spin angular momentum operators, $\expec{S_x}=\expec{S_y}=\expec{S_z}=0$, for all the three approximate eigenstates $\ket{0}$ and $\ket{\pm}$. The energy levels become insensitive to the Zeeman and hyperfine interactions in the first-order \cite{Genovese_TempSens,Yacoby_AngleSens2021npj}. This is known as ZEFOZ shift \cite{ZEFOZsellars1}. The effect of the spin bath on the NV electron spin can be seen as a fluctuating magnetic noise that shifts the transition frequencies. However, at ZEFOZ points, the gradient of the transition frequencies with respect to the magnetic field is strongly suppressed. Due to this, spin systems tend to have long coherence times at ZEFOZ points \cite{ZEFOZsellars1,ZEFOZsuter,Koti2020_LAC}. 

\begin{figure}
	\centering \includegraphics[width=8.5cm]{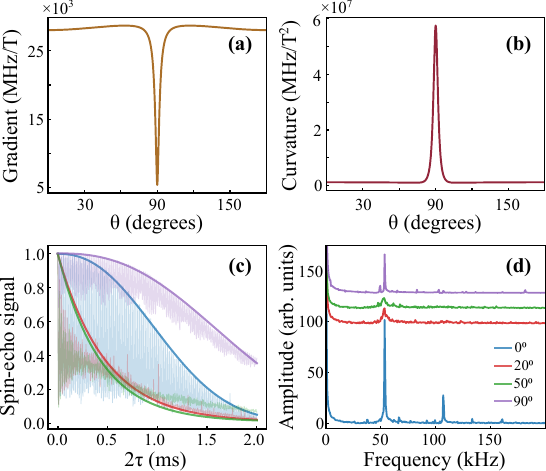} 
    \caption{(a) and (b) Calculated gradient and curvature of an electron spin transition frequency with respect to magnetic field noise as a function of $\theta$ for $B=10$ mT. (c) and (d) CCE-2 spin-echo simulation results of the electron spin transition $\ket{0}\leftrightarrow\ket{-1}$ (or $\ket{-}$) for different values of $\theta$ and their Fourier transforms (FFTs). The $T_2$ values are 1.2, 0.5, 0.4, and 2.0 ms for $\theta=0^\circ$, $20^\circ$, $50^\circ$, and $90^\circ$ respectively.}
	\label{fig:Gradient} 
\end{figure}

For parallel field, the gradient of the transition frequencies with respect to magnetic field is equal to $\gamma_e$. At transverse field, it is approximately equal to $2\gamma_e^2 B/D$ (Supplementary Section S7).
Though the gradient is not exactly zero here, it is significantly lower for field strengths satisfying $\gamma_e B\ll D$.
As the first-order magnetic-field sensitivity is suppressed, the contribution of second-order magnetic-field fluctuations to decoherence becomes more significant.
This effect can be quantified by the curvature (second derivative) of the transition frequencies with respect to the magnetic field \cite{Basanta2025hBN}. Gradient and curvature of the $\ket{0}\leftrightarrow\ket{-}$ transition as a function of $\theta$ are shown in Figs. \ref{fig:Gradient}(a) and (b). The gradient decreases sharply at $\theta=90^\circ$, while the curvature increases rapidly. Though the curvature is high at $\theta=90^\circ$, its contribution to decoherence is orders of magnitude smaller than that of the gradient. Hence, we expect an improvement in $T_2$ for a transverse field.

In order to quantify the dependence of coherence time on $\theta$, the spin-bath dynamics are modeled using the cluster correlation expansion (CCE) method \cite{Witzel2006CCE,Yang2008CCE} (Supplementary Section S5).
The results plotted in Figs. \ref{fig:Gradient}(c) and (d) show that $T_2$ decreases as $\theta$ increases, but increases again at $\theta=90^{\circ}$, where it is nearly twice its value at $\theta=0^\circ$, in agreement with the experiment.
Nevertheless, at $\theta=90^\circ$, CCE-2 does not reproduce the observed ESEEM. Instead, it shows low-amplitude modulations at $\frac{1}{2}\gamma_C B$. The nuclear spin of the host $^{14}$N atom is excluded from the CCE-2 simulations, and it appears that the observed modulation frequency is related to it.

\begin{figure}
	\centering \includegraphics[width=8.8cm]{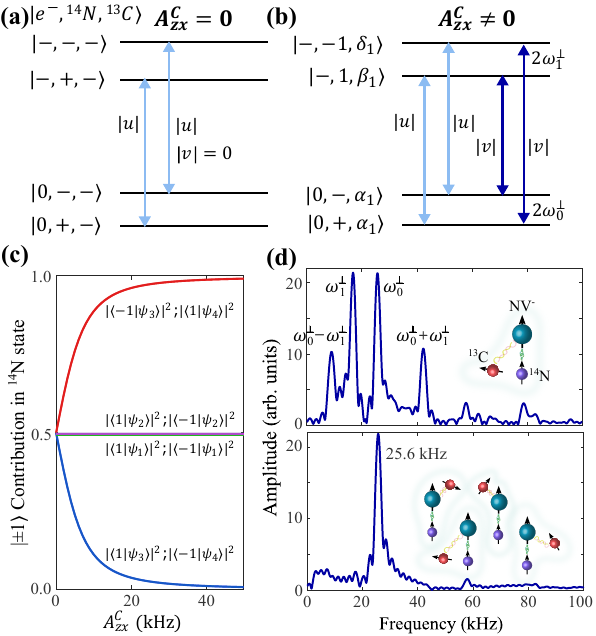} 
	\caption{Electron spin-echo simulations of a $^{14}$N-NV-$^{13}$C three-spin system in a transverse magnetic field. Relevant energy levels for (a) the $^{13}$C hyperfine parameter $A_{zx}^C=0$ and (b) $A_{zx}^C=0.25$ MHz. (c) Change in $^{14}$N eigenstates as a function of $A_{zx}^C$. $\ket{\psi_{1,2}}$ are $^{14}$N states in the $\ket{0}$ electron spin manifold, while $\ket{\psi_{3,4}}$ belong to the $\ket{-}$ manifold. (d) Fourier transforms of simulated spin-echo signals at 10.7 mT. The upper spectrum is simulated with the $^{13}$C parameters $[A_{xx},A_{yy},A_{zz},A_{xy},A_{zx},A_{zy}]= [0.17,0.08,0.13,0.10,0.25,0.05]$ MHz; the lower one is averaged over 11 $^{13}$C hyperfine couplings with $A_{zx}^C=0.05-0.25$ MHz. For $A_{zx}^C=0.25$, $\ket{\alpha_1}\approx 0.81\ket{\frac{1}{2}}+0.58\ket{-\frac{1}{2}}$, $\ket{\beta_1}\approx 0.74\ket{\frac{1}{2}}+0.66\ket{-\frac{1}{2}}$, $\ket{\delta_1}\approx 0.63\ket{\frac{1}{2}}+0.78\ket{-\frac{1}{2}}$.}
	\label{fig:Sim} 
\end{figure}

The ESEEM oscillations observed in the experiment at the transverse field can be explained by using a 3-spin system containing the NV electron spin, $^{14}$N nuclear spin, and a $^{13}$C nuclear spin of the bath. The corresponding Hamiltonian of the system can be written as
\begin{align}
	&{\cal H}_2 = \ DS_z^2 - \gamma_e \vec{B}\cdot\vec{S}+A_\parallel^N S_zI_z^N +A_\perp^N (S_xI_x^N+S_yI_y^N) 
	\nonumber\\
	&+P (I_z^N)^2 - \gamma_N \vec{B}\cdot\vec{I}^N +\vec{S}\cdot\bm{A^C}\cdot\vec{I}^C - \gamma_C \vec{B}\cdot\vec{I}^C,
\end{align}
where $A_\parallel^N$ and $A_\perp^N$ are the parallel and perpendicular components of the $^{14}$N hyperfine interaction. $P=-4.95$ MHz and $\gamma_N=3.08$ MHz/T are the quadrupole coupling and gyromagnetic ratio of the $^{14}$N nuclear spin respectively. $\bm{A^C}$ represents the hyperfine tensor of the $^{13}$C nuclear spin. Using this Hamiltonian, we simulate the electron spin-echo signal of the $\ket{0}\leftrightarrow\ket{-}$ transition (Supplementary Section S5). 
The main features of the simulation can be explained using four of the twelve energy levels, two from each of the $\ket{0}$ and $\ket{-}$ subspaces, shown in Fig. \ref{fig:Sim}.
The eigenstates are ordered as $\ket{e^-,{^{14}N},{^{13}C}}$. For the $^{14}$N nucleus, $\ket{\pm}=\frac{\ket{-1}\pm \ket{1}}{\sqrt{2}}$ and for $^{13}$C, $\ket{\pm}=\frac{1}{\sqrt{2}}\left(\ket{-\frac{1}{2}}\pm \ket{\frac{1}{2}}\right)$. The complete energy level diagram is given in Supplementary Section S4.


The term $A_{zx}^C S_zI_x$ of the $^{13}$C hyperfine interaction is crucial in explaining the experimental echo modulations. Here $z$ and $x$ refer to the NV-axis and the direction of the magnetic field perpendicular to the NV-axis respectively. The energy level diagram for the case $\theta=90^\circ$ and $A_{zx}^C=0$ is shown in Fig. \ref{fig:Sim}(a). Here, the nuclear spin states are independent of the electron spin state. The allowed electron spin transitions are $\ket{0,+,+}\leftrightarrow\ket{-,+,+}$ and $\ket{0,-,+}\leftrightarrow\ket{-,-,+}$, which have the same nuclear spin states at both ends. The first $\pi/2$ pulse of the spin-echo sequence excites these transitions. However, the spin-echo signal does not exhibit any modulation.
The energy level diagram for $A_{zx}^C=0.25$ MHz is shown in Fig. \ref{fig:Sim}(b). Here, if the electron spin state is $\ket{0}$, the spin states of the $^{14}$N nucleus are $\ket{\pm}$, and if the electron spin state is $\ket{-}$, the $^{14}N$ spin states are $\ket{\pm 1}$, i.e., \textit{the $^{14}$N nuclear spin state is conditioned on the electron spin state because of an off-diagonal element of the $^{13}$C hyperfine tensor}.
The change in $^{14}$N nuclear spin states as a function of the hyperfine coupling of a bath $^{13}$C nuclear spin with the electron spin is shown in Fig. \ref{fig:Sim}(c). The states ($\ket{\psi_1}$ and $\ket{\psi_2}$) that belong to the $\ket{0}$ electron spin manifold remain $\ket{\pm}$ irrespective of the $^{13}$C hyperfine coupling strength. However, the states ($\ket{\psi_3}$ and $\ket{\psi_4}$) that belong to the $\ket{-}$ electron spin manifold change rapidly from $\ket{\pm}$ to $\ket{\pm 1}$ as the parameter $A_{zx}^C$ changes from 0 to 20 kHz. These $^{14}$N states change significantly even for a few kHz of $A_{zx}^C$ value.


All the four electron spin transitions between the energy levels shown in Fig. \ref{fig:Sim}(b) are allowed. The first $\pi/2$-pulse of the spin-echo sequence simultaneously excites all the four transitions. Quantum interference between them creates quantum beats at the nuclear spin transition frequencies $\ket{0,+,\alpha_1}\leftrightarrow\ket{0,-,\alpha_1}$ and $\ket{-,-1,\beta_1}\leftrightarrow\ket{-,1,\delta_1}$ \cite{QBeatsForrester1955,QBeatsHadeishi1965,QBeatsHaroche1973}. In this scenario, the spin-echo signal can be written as \cite{Mims_ESEEM1972}
\begin{align}
	{\cal S}(t) = &|u|^4 + |v|^4 + |u|^2 |v|^2 [2 \cos(\omega_0^\perp t) + 2 \cos(\omega_1^\perp t) \nonumber \\
	& - \cos((\omega_0^\perp-\omega_1^\perp)t) - \cos((\omega_0^\perp+\omega_1^\perp)t)],
	\label{eq:echo}
\end{align}
where $|u|$ and $|v|$ represent the electron-spin transition amplitudes, and  $2\omega_0^\perp$ and $2\omega_1^\perp$ are the frequency differences between the nuclear-spin sublevels corresponding to different electron-spin states as shown in Fig. \ref{fig:Sim}(b).

The Fourier transformed spectrum of the simulated spin-echo signal for a transverse field of 10.7 mT and $A_{zx}^C=0.25$ MHz is shown in Fig. \ref{fig:Sim}(d) (upper trace). The peak frequencies and their relative amplitudes are in excellent agreement with Eq. \ref{eq:echo}. 
The frequency $\omega_0^\perp$ depends on $A_\perp^N$, $P$, and $B$, whereas $\omega_1^\perp$ depends on $A_\parallel^N$, $A_{zx}^C$, and $B$.
Consequently, the peaks at $\omega_1^\perp$, $\omega_0^\perp-\omega_1^\perp$, and $\omega_0^\perp+\omega_1^\perp$ average out when averaged over different $^{13}$C hyperfine coupling strengths (0.05 - 0.25 MHz) as shown in the lower trace of Fig. \ref{fig:Sim}(d).
$\omega_0^\perp$, which is equal to one-half of the $^{14}$N transition frequency $\ket{0,+,\alpha_1}\leftrightarrow\ket{0,-,\alpha_1}$, does not depend on $^{13}$C hyperfine parameters. Correspondingly the peak at $\omega_0^\perp$ survives the averaging.
Its frequency (25.6 kHz) exactly matches the experimental ESEEM frequency at 10.7 mT for $P=-4.95$ MHz and $A_\perp^N=-2.64$ MHz. Furthermore, the plot of $\omega_0^\perp$ as a function of $B$ in Fig. \ref{fig:FreqVsField} fits very well with all the measured modulation frequencies. 
The spin-echo modulation at $\omega_0^\perp$ occurs even when multiple $^{13}$C nuclear spins are coupled to the NV center.
The results of a simulation consisting of three $^{13}$C nuclear spins coupled to an NV center are shown in Supplementary Section S6.


The observed spin-echo modulation frequency ($\omega_0^\perp$) arises from second-order interactions at the transverse magnetic field, involving the $^{14}$N nuclear spin parameters $P$ and $A_\perp^N$.
These parameters are less sensitive to temperature and strain than the NV electron spin parameters $D$ and $E$ \cite{Jarmola2020tempQ, Lourette2023Temp, Liu2023DFTStressHyperfine}, making the modulation frequency robust against thermal and strain fluctuations.
The slope of the modulation frequency with respect to the transverse field at 10 mT is 4.44 MHz/T, which is much smaller than the typical electron spin gradient $\gamma_e=28025$ MHz/T. The long coherence times of the modulations can partly compensate their low magnetic-field sensitivity. Thus, the observed ESEEM frequency is highly stable against environmental variations.

In conclusion, we have demonstrated spin-bath mediated, long-lived electron spin-echo envelope modulations of NV centers in diamond in transverse magnetic fields. The long coherence times can be attributed to the ZEFOZ shift at the energy level anti-crossing point induced by the transverse field. The modulation frequency corresponds to a nuclear spin transition of the host $^{14}$N atom. However, the electron spin-echo envelope is modulated by this frequency because the $^{13}$C nuclear spin bath modifies the eigenstates of the electron-$^{14}$N nuclear spin system.
These findings show that the role of the spin bath in solid-state spin systems extends beyond causing decoherence and dissipation, as it can also modify the eigenstates of the system, thereby blurring the conventional distinction between the system and its environment.
This work opens new directions for understanding the role of quantum many-body interactions in solid-state spin systems. Although demonstrated here using NV centers in diamond, the underlying mechanism is expected to occur in similar solid-state defects coupled to nuclear spin baths, e.g., NV centers in silicon carbide.

\section*{Acknowledgements}
We thank T. S. Mahesh for the generous loan of one of the diamond samples used in this work. A.R. and R.K.K. acknowledge support from Department of Science \& Technology - Science \& Engineering Research Board (DST-SERB), India through grant no. SRG/2020/000765. 
S.D. thanks the Indian Institute of Technology, Madras, India, and the Science and Engineering Research Board (SERB Grant No. SRG/2023/000322), India, for start-up funding. S.D. and B.M. acknowledge the use of the computational facilities supported by a grant from the Mphasis F1 Foundation given to the Center for Quantum Information, Communication, and Computing (CQuICC).

\bibliography{bibNV26}

\end{document}


\title{Supplementary Information:\\
    Spin bath mediated long-lived coherent oscillations of NV centers in diamond}
    \date{\vspace{-3em}}
    \maketitle
    \onehalfspacing

\renewcommand{\thesection}{S\arabic{section}}
\setcounter{section}{0}

\section{Experimental details}

\subsection{Samples}
Two chemical vapor deposition grown single crystal diamond samples of dimensions $3 \times 3 \times 0.5$ mm and $2 \times 2 \times 1$ mm are used in this work. Their nitrogen concentration is about 1 ppm and they contain natural abundance of $^{13}$C atoms.

\subsection{Setup}
All the experiments of this work have been performed on a home-built confocal microscope setup equipped with microwave and radio-frequency electronics for the resonant excitation of the electron spins of NV centers. A 532 nm diode pumped solid-state laser is used to initialize and readout the electron spins. A single-photon detector based on an avalanche photo-diode is used to measure the fluorescence of NV centers. A 560 nm long-pass dichroic mirror separates the laser and  fluorescence beam paths. The laser light, reflected by the dichroic mirror, is focused onto the diamond crystal by using an objective lens having 1.45 NA. The fluorescence emitted by the NV centers is collected by the same objective lens and it gets transmitted through the dichroic mirror. A 50 $\mu$m pinhole is used in the detection path to filter the out of plane fluorescence. The lateral and axial resolutions of the confocal spot are about 0.25 and 1 $\mu$m respectively. From the nitrogen concentration of the diamond crystal and the photoluminescence intensity, we can estimate that there are hundreds of NV centers in the confocal volume. Static magnetic field is applied to NV centers by using a permanent magnet. The orientation of the static field with respect to the diamond crystal is controlled by using two rotation stages connected to the magnet.

\section{$T_2$ for different $\theta$ values}

\begin{figure}
	\centering \includegraphics[width=8cm]{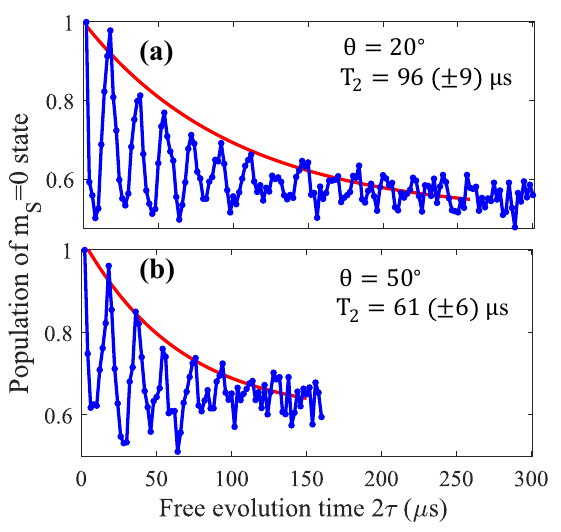} 
	\caption{Spin-echo signals of ensemble NV centers  measured in a static magnetic field of 10.7 mT oriented at (a) $20^\circ$ and (b) $50^\circ$ with respect to the NV-axis. Red curves correspond to the decay of the echo revivals fit to the equation $a\exp(-2\tau/T_2)^n+b$. The revival frequency is equal to $\frac{1}{2} \gamma_C B$. Errors in $T_2$ correspond to one standard deviation confidence intervals obtained from the fit.}
	\label{fig:EchoSig} 
\end{figure}

It has been demonstrated in Ref. \cite{WalsworthDecoh} that the spin-echo coherence time ($T_2$) of NV centers decreases as the angle ($\theta$) between the NV axis and the static magnetic field increases. We also observe this experimentally as shown in Fig. \ref{fig:EchoSig}. At $\theta=20^\circ$, $T_2=96$ ($\pm 9$) $\mu$s and at $\theta=50^\circ$, $T_2=61$ ($\pm 6$) $\mu$s, which are much smaller compared to $T_2=200$ $\mu$s at $\theta=0^\circ$. $T_2$ continues to decrease till around $\theta=84^\circ$. However, from $\theta=85^\circ$ to $90^\circ$, $T_2$ increases again as shown in Fig. 2 of the main text.

\section{Electron spin-echo of $\ket{-}\leftrightarrow\ket{+}$ transition}

\begin{figure}
	\centering \includegraphics[width=10cm]{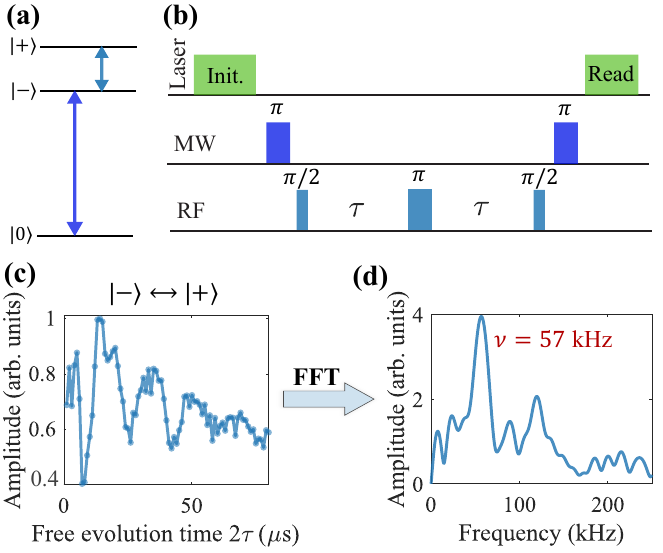} 
	\caption{Spin-echo measurement of the electron spin $\ket{-}\leftrightarrow\ket{+}$ transition in a transverse magnetic field. (a) Energy level diagram. (b) Pulse sequence. Microwave (MW) $\pi$-pulses are applied on resonance with the $\ket{0}\leftrightarrow\ket{-}$ transition, and radio-frequency (RF) spin-echo pulses are applied on resonance with the $\ket{-}\leftrightarrow\ket{+}$ electron spin transition, whose frequency is 32 MHz at a transverse field of 10.7 mT. (c) Time-domain spin-echo signal and (d) its Fourier transform.}
	\label{fig:EchoForbid} 
\end{figure}

For a transverse magnetic field, the transition between the electron spin states $\ket{-}$ and $\ket{+}$ is allowed as illustrated in Fig. \ref{fig:EchoForbid}(a). The spin-echo signal of this transition is measured by using the pulse sequence shown in Fig. \ref{fig:EchoForbid}(b). Since optical pumping does not directly create a population difference between the states $\ket{-}$ and $\ket{+}$, microwave (MW) $\pi$-pulses on resonance with the transition $\ket{0}\leftrightarrow\ket{-}$ are used to transfer population from the $\ket{0}$ state to the $\ket{-}$ state and back (for readout). The radio-frequency (RF) pulses, shown in the third row of Fig. \ref{fig:EchoForbid}(b), are applied on resonance with the electron spin transition $\ket{-}\leftrightarrow\ket{+}$, whose transition frequency is 32 MHz in a transverse magnetic field of 10.7 mT. The resulting spin-echo signal and its Fourier transform are shown in Figs. \ref{fig:EchoForbid}(c) and (d) respectively. The spin-echo signal exhibits decays and revivals at one-half of the $^{13}$C Larmor precession frequency ($\frac{1}{2} \gamma_C B=57$ kHz) in contrast to the spin-echo envelope modulations observed for the $\ket{0}\leftrightarrow\ket{-}$ and $\ket{0}\leftrightarrow\ket{+}$ transitions in a transverse field. However, the coherence time $T_2$ of this transition is significantly shorter than those of the transitions $\ket{0}\leftrightarrow\ket{-}$ and $\ket{0}\leftrightarrow\ket{+}$.

\section{Energy level diagram of the NV center in a transverse magnetic field}

\begin{figure}
	\centering \includegraphics[width=10cm]{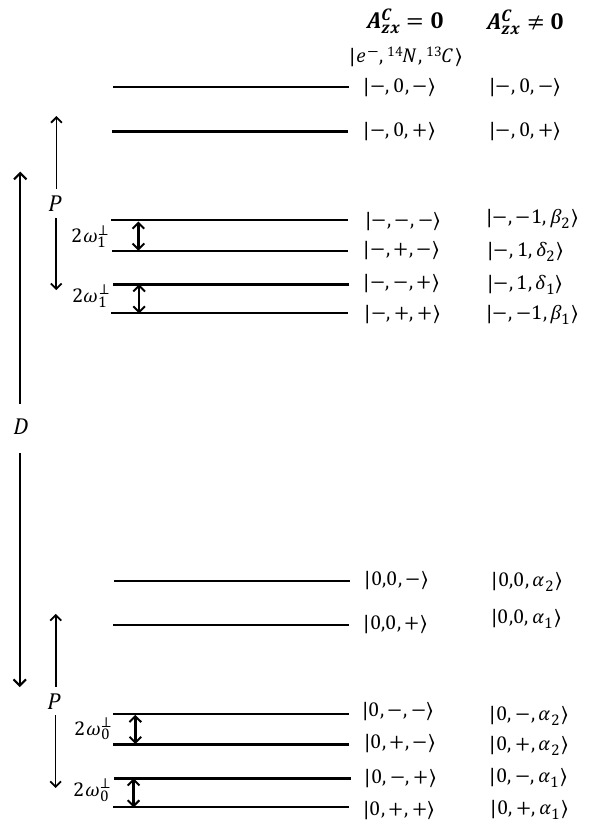} 
	\caption{Energy level diagram corresponding to the $\ket{0}$ and $\ket{-}$ states of the NV electron spin coupled to the $^{14}$N nuclear spin and a $^{13}$C nuclear spin, in a transverse magnetic field of 10.7 mT. The eigenstates are given for both without and with the $^{13}$C hyperfine coupling term $A_{zx}^C S_zI_x$ in the Hamiltonian. For the electron and $^{14}$N nuclear spins $\ket{\pm}=\frac{\ket{-1}\pm \ket{1}}{\sqrt{2}}$ and for the $^{13}$C $\ket{\pm}=\frac{1}{\sqrt{2}}\left(\ket{-\frac{1}{2}}\pm \ket{\frac{1}{2}}\right)$. For $A_{zx}^C=0.25$ MHz, $\ket{\alpha_1}\approx 0.81\ket{\frac{1}{2}}+0.58\ket{-\frac{1}{2}}$, $\ket{\beta_1}\approx 0.74\ket{\frac{1}{2}}+0.66\ket{-\frac{1}{2}}$, $\ket{\delta_1}\approx 0.63\ket{\frac{1}{2}}+0.78\ket{-\frac{1}{2}}$. $\ket{\alpha_2}$, $\ket{\beta_2}$, and $\ket{\delta_2}$ are states orthogonal to $\ket{\alpha_1}$, $\ket{\beta_1}$, and $\ket{\delta_1}$ respectively.}
	\label{fig:Englvl} 
\end{figure}

The energy level diagram of a three-spin system consisting of the NV electron, $^{14}$N nucleus and a bath $^{13}$C nucleus, relevant for the electron spin transition $\ket{0}\leftrightarrow\ket{-}$, in a transverse magnetic field of 10.7 mT is shown in Fig. \ref{fig:Englvl}. The spin states are arranged in the order $\ket{e^-,{^{14}N},{^{13}C}}$.
The eigenstates of the $^{14}$N nuclear spin depend on the $^{13}$C nuclear spin hyperfine coupling term $A_{zx}^C S_zI_x$. If $A_{zx}^C=0$, the $^{14}$N nuclear spin eigenstates are $\ket{0}$ and $\ket{\pm}$ for both the $\ket{0}$ and $\ket{-}$ subspaces of the electron spin. However, for $A_{zx}^C\neq0$, the $^{14}$N eigenstates are $\ket{0}$ and $\ket{\pm}$ if the electron spin is in the $\ket{0}$ state, and they are $\ket{0}$ and $\ket{\pm 1}$ if the electron spin is in the $\ket{-}$ state. That is, the $^{14}$N nuclear spin state is conditioned on the electron spin state because of the hyperfine coupling of the $^{13}$C nuclear spin.

\section{Simulation details}

\subsection{CCE-simulation}
Here, we consider that the central NV electron spin ($S=1$) is interacting with a surrounding bath of $^{13}$C ($I=1/2$) nuclear spins.
The $^{13}$C spin bath is generated by randomly occupying sites of the diamond lattice according to the natural isotopic abundance ($\sim 1.1\%$). Interactions between the NV electronic spin and the nuclear spins are modeled using the purely dipolar hyperfine tensor. For each realization, the bath radius is fixed to $3~\mathrm{nm}$, while the maximum separation between correlated $^{13}$C pairs included in the calculation is restricted to $0.5~\mathrm{nm}$. These parameters resulted in bath realizations containing approximately 220 individual $^{13}$C spins and about 100 coupled $^{13}$C pairs.
The Hahn-echo coherence signal is evaluated within the CCE-2 approximation by including both single-spin and pair-cluster contributions \cite{Yao2006CCE}. To account for statistical variations in the spin environment and, thus, to mimic ensemble effects, the simulations are averaged over 20 independent bath realizations, and the resulting averaged coherence is presented in Fig. 4(c) of the main text. The detailed CCE workflow is as follows.

For a given magnetic field, the NV electron spin Hamiltonian is diagonalized and the two lowest-energy eigenstates are selected to define an effective qubit. The spin operators are projected onto this dressed-state subspace according to
\begin{equation}
	\tilde{S}_{\alpha}
	=
	P S_{\alpha} P,
\end{equation}
where $P$ is the projector onto the chosen two-level subspace.

The hyperfine interaction tensor between the NV electron spin and the $i$th $^{13}$C nuclear spin is approximated by the magnetic dipole--dipole interaction,
\begin{equation}
	A_{i,\alpha\beta}
	=
	\frac{\mu_0}{4\pi}
	\frac{\gamma_e\gamma_C\hbar}{r_i^3}
	\left(
	\delta_{\alpha\beta}
	-
	3\hat r_{i,\alpha}\hat r_{i,\beta}
	\right),
	\qquad
	\alpha,\beta\in\{x,y,z\},
\end{equation}
where $r_i=|\vec{r}_i|$ is the distance between the NV center and the $i$th nucleus, $\hat{r}_i=\vec{r}_i/r_i$ is the corresponding unit vector, $\hat r_{i,\alpha}$ denotes its $\alpha$th Cartesian component, and $\delta_{\alpha\beta}$ is the Kronecker delta.

Let $\{|0\rangle,|1\rangle\}$ denote the two dressed NV qubit states. The effective Hamiltonian of a single nuclear spin conditioned on the NV qubit state $m\in\{0,1\}$ is
\begin{equation}
	H_i^{(m)}
	=
	\gamma_C \vec{B} \cdot \vec{I}_i
	+
	\sum_{\alpha,\beta}
	\langle m|\tilde S_{\alpha}|m\rangle
	A_{i,\alpha\beta}
	I_{i,\beta}.
\end{equation}
To incorporate pairwise bath correlations, pairs of $^{13}$C nuclei separated by less than $0.5~\mathrm{nm}$ are retained. 
The effective Hamiltonian of a nuclear-spin pair $(i,j)$ is
\begin{equation}
	H_{ij}^{(m)}
	=
	H_i^{(m)}
	+
	H_j^{(m)}
	+
	H_{ij}^{\mathrm{dip}},
\end{equation}
where the nuclear dipolar interaction is
\begin{equation}
	H_{ij}^{\mathrm{dip}}
	=
	\frac{\mu_0}{4\pi}
	\frac{\gamma_C^2\hbar}{r_{ij}^3}
	\left[
	\vec{I}_i\cdot\vec{I}_j
	-
	3
	(\vec{I}_i\cdot\hat{r}_{ij})
	(\vec{I}_j\cdot\hat{r}_{ij})
	\right].
\end{equation}

\noindent \textit{Hahn Echo Evolution:}
For each cluster, the conditional propagators associated with the two dressed NV qubit states are
\begin{equation}
	U_C^{(m)}(\tau)
	=
	e^{-iH_C^{(m)}\tau},
\end{equation}
where $m\in{0,1}$ labels the two dressed NV qubit states. The coherence contribution of a cluster is calculated as
\begin{equation}
	L_C(2\tau)
	=
	\mathrm{Tr}
	\left[
	U_C^{(1)\dagger}
	U_C^{(0)\dagger}
	U_C^{(1)}
	U_C^{(0)}
	\rho_C
	\right],
\end{equation}
where $\rho_C$ denotes the maximally mixed initial state of the cluster.
The single-spin coherence contribution of the $i$th nucleus is denoted by $L_i(2\tau)$, while the pair-cluster coherence of nuclei $(i,j)$ is denoted by $L_{ij}(2\tau)$. To avoid double counting of independent single-spin dynamics, the irreducible pair contribution is defined as
\begin{equation}
	\tilde{L}_{ij}(2\tau)
	=
	\frac{L_{ij}(2\tau)}
	{L_i(2\tau)L_j(2\tau)}.
\end{equation}
The total coherence within the CCE-2 approximation is then given by
\begin{equation}
	L(2\tau)
	=
	\prod_i L_i(2\tau)
	\prod_{i<j}
	\tilde{L}_{ij}(2\tau).
\end{equation}

\subsection{Electron spin-echo simulation}
The Hamiltonian of the three spin system consisting of the NV electron spin, $^{14}$N nuclear spin, and a $^{13}$C nuclear spin is given by ${\cal H}_2$ in Eq. 1 of the main text. The Hamiltonian corresponding to the microwave field can be written as
\begin{equation}
	{\cal H}_{\textrm{mw}} (t) = \gamma_e B_{\textrm{mw}} (\cos \eta\ S_x + \sin \eta\ S_y) \cos(\omega t +\varphi),
\end{equation}
where $\eta$ is the angle between the directions of the static and microwave magnetic fields in the transverse plane. $B_{\textrm{mw}}$, $\omega$, and $\varphi$ represent the amplitude, frequency, and phase of the microwave field respectively.

The unitary operator corresponding to the microwave pulses of the spin-echo sequence can be represented as 
\begin{equation}
	U_{p}= {\cal{T}} \exp \left(-i \int_0^t ({\cal H}_2 + {\cal H}_{\textrm{mw}} (t'))\ dt'\right),
\end{equation}
where $\cal{T}$ is the Dyson time-ordering operator and $t$ is the duration of the pulse. The unitary operator is evaluated by discretizing the pulse duration $t$ into $N$ intervals of duration $\Delta t=\frac{t}{N}$. Here $N$ is chosen such that $\Delta t \ll \frac{1}{\|{\cal H}_2\|}$.
So, the unitary operator can be approximated as 
\begin{equation}
	U_{p} \approx U_N U_{N-1} \cdots U_n \cdots U_2 U_1,
\end{equation}
where
\begin{equation}
	U_n = \exp \left(-i \int_{(n-1)\Delta t}^{n \Delta t} ({\cal H}_2 + {\cal H}_{\textrm{mw}}(t'))\ dt'\right).
\end{equation}
The unitary operator corresponding to the free-evolution time $\tau$ can simply be written as 
\begin{equation}
	U_\tau = \exp \left(-i{\cal H}_2 \tau\right).
\end{equation}

The electron spin is initialized into the $\ket{0}$ state, and the $^{14}$N and $^{13}$C nuclear spins are in thermal equilibrium states, which are close to maximally mixed states. The corresponding initial density matrix of the spin system can be written as $\rho_i=\ket{0}\bra{0}\otimes \frac{1}{3}\mathbbm{1}_3\otimes \frac{1}{2}\mathbbm{1}_2$, where $\mathbbm{1}_2$ and  $\mathbbm{1}_3$ represent $2\times 2$ and $3\times 3$ identity matrices respectively. 
The final density matrix of the system after applying the spin-echo pulse sequence is
\begin{equation}
	\rho_f = U_{\pi/2} U_\tau U_{\pi} U_\tau U_{\pi/2} (\rho_i) U_{\pi/2}^{\dagger} U_{\tau}^{\dagger} U_{\pi}^{\dagger}U_{\tau}^{\dagger} U_{\pi/2}^{\dagger},
\end{equation}
where $U_{\pi/2}$ and $U_{\pi}$ represent unitary operators of $\pi/2$ and $\pi$ pulses respectively.
The population of the electron spin $\ket{0}$ state is calculated by using the expression 
\begin{equation}
	P_{\ket{0}} = tr(\rho_f \ket{0} \bra{0}).
\end{equation}
The time-domain spin-echo signal is calculated as ${\cal S}(\tau)= P_{\ket{0}} \exp (-2\tau/T_2)$, where the exponential factor accounts for the decay of the signal. The spectrum is obtained by Fourier transforming ${\cal S}(\tau)$. 

\section{Simulation of electron spin-echo for a 5-spin system}
\begin{figure}
	\centering \includegraphics[width=16cm]{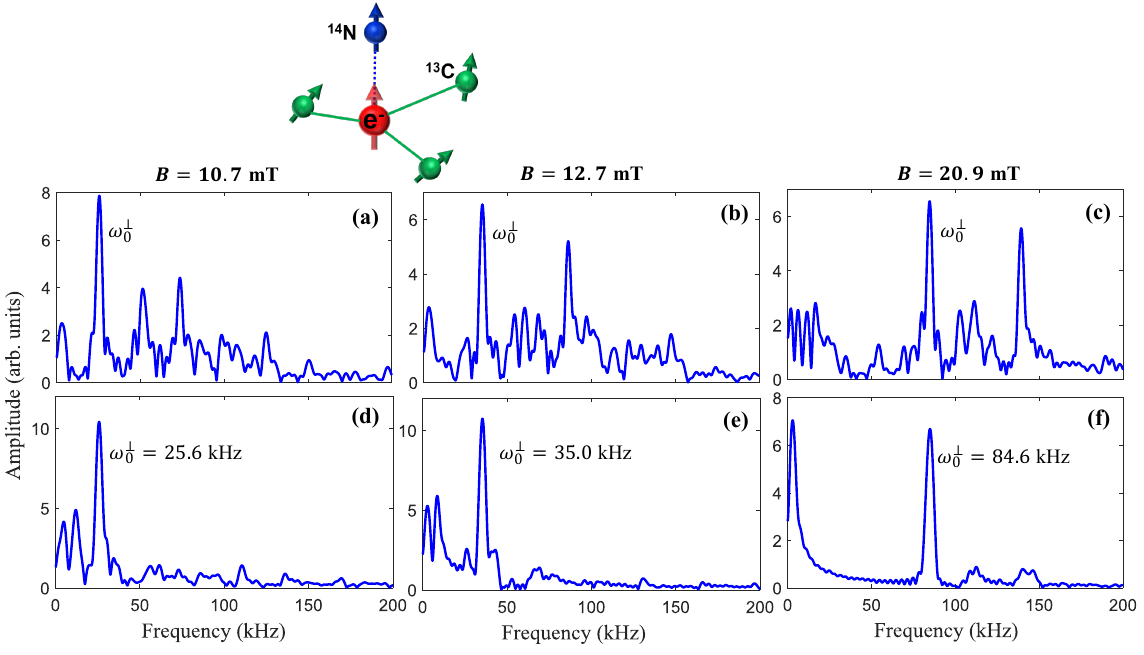} 
	\caption{Fourier transforms of the simulated electron spin-echo signals of a five-spin system (NV electron spin, $^{14}$N nuclear spin, and three $^{13}$C nuclear spins) for three different amplitudes of the transverse magnetic field and two different sets of $^{13}$C hyperfine tensors. (a), (b) and (c) are simulated using the hyperfine tensors (expressed in MHz) of Eq. \ref{eq:hyptens1}, and (d), (e) and (f) are simulated using those of Eq. \ref{eq:hyptens2}.}
	\label{fig:5spinsiml} 
\end{figure}

In the main text, electron spin-echo simulations of a 3-spin system consisting of the NV electron spin, the $^{14}$N nuclear spin, and a $^{13}$C nuclear spin are shown. Here, in Fig. \ref{fig:5spinsiml}, we show the results of simulations performed on a 5-spin system comprising an NV center coupled to three $^{13}$C nuclear spins. Spectra shown in Figs. \ref{fig:5spinsiml}(a)--(c) are simulated using the $^{13}$C hyperfine tensors (expressed in MHz) given in Eq. \ref{eq:hyptens1}, for the transverse field strengths 10.7 mT, 12.7 mT, and 20.9 mT, respectively. Those shown in Figs. \ref{fig:5spinsiml}(d)--(f) are simulated using the tensors in Eq. \ref{eq:hyptens2} for the same fields. 
The spectra show several peaks. However, the most prominent of them is the peak at the frequency $\omega_0^\perp$, which corresponds to the $^{14}$N transition $\ket{+}\leftrightarrow\ket{-}$ within the $\ket{0}$ subspace of the electron spin.
This peak appears in the spectrum if there is at least one $^{13}$C nuclear spin coupled to the electron spin with the hyperfine parameter $A_{zx}^C$ having a value of 10 kHz or higher. It is independent of the values of the other hyperfine parameters or the number of $^{13}$C nuclear spins coupled. The frequencies and amplitudes of other peaks seen in the spectra depend on the $^{13}$C hyperfine coupling strengths. As shown in Fig. \ref{fig:5spinsiml}, the frequency $\omega_0^\perp$ increases non-linearly with the magnetic field strength, and its values observed in the 5-spin simulations match very well with those in the 3-spin simulation and the experimental spin-echo envelope modulation frequencies shown in Fig. 3 of the main text.

\begin{align}
	\bm{A_1^C} = 
	\begin{pmatrix}
		-0.063 & -0.223 & 0.461 \\
		-0.223 & -0.320 & -0.266 \\
		0.461 & -0.266 & 0.611
	\end{pmatrix},
	\bm{A_2^C} =
	\begin{pmatrix}
		0.620 & -0.028 & 0.280 \\
		-0.028 & 0.442 & 0.097 \\
		0.280 & 0.097 & 0.314
	\end{pmatrix}, \nonumber\\
	\bm{A_3^C}=
	\begin{pmatrix}
		0.039 & -0.020 & 0.140 \\
		-0.020 & -0.059 & -0.027 \\
		0.140 & -0.027 & 0.113
	\end{pmatrix}.
	\label{eq:hyptens1}
\end{align}

\begin{align}
	\bm{A_1^C} = 
	\begin{pmatrix}
		-0.028 & -0.014 & -0.112 \\
		-0.014 & -0.106 & 0.015 \\
		-0.112 & 0.015 & 0.052
	\end{pmatrix},
	\bm{A_2^C} =
	\begin{pmatrix}
		0.078 & 0.059 & -0.013\\
		0.059 & -0.052 & -0.004 \\
		-0.013 & -0.004 & -0.094
	\end{pmatrix}, \nonumber \\
	\bm{A_3^C}=
	\begin{pmatrix}
		0.510 & 0.090 & -0.056 \\
		0.090 & 0.551 & -0.291 \\
		-0.056 & -0.291 & 0.314
	\end{pmatrix}.
		\label{eq:hyptens2}
\end{align}

\section{Gradient and Curvature Calculations}

The spin Hamiltonian is
\begin{equation}
	H = D S_z^2 + \gamma_e B_x S_x + \gamma_e B_y S_y + \gamma_e B_z S_z
	+ \gamma_e ( b_x S_x + b_y S_y + b_z S_z ),
\end{equation}
where $ \vec{b} = (b_x,b_y,b_z)$ denotes the fluctuating magnetic field. Defining the total magnetic field components
\begin{equation}
	\bar{B}_x = B_x + b_x, \qquad
	\bar{B}_y = B_y + b_y, \qquad
	\bar{B}_z = B_z + b_z ,
\end{equation}
the Hamiltonian becomes
\begin{equation}
	H = D S_z^2 + \gamma_e \bar{B}_x S_x
	+ \gamma_e \bar{B}_y S_y
	+ \gamma_e \bar{B}_z S_z .
\end{equation}
In the $\{ \ket{+1}, \ket{0}, \ket{-1} \}$ basis, the Hamiltonian matrix reads
\begin{equation}
	H =
	\begin{pmatrix}
		D + \gamma_e \bar{B}_z
		& \dfrac{\gamma_e}{\sqrt{2}}(\bar{B}_x - i \bar{B}_y)
		& 0 \\[8pt]
		\dfrac{\gamma_e}{\sqrt{2}}(\bar{B}_x + i \bar{B}_y)
		& 0
		& \dfrac{\gamma_e}{\sqrt{2}}(\bar{B}_x - i \bar{B}_y) \\[8pt]
		0
		& \dfrac{\gamma_e}{\sqrt{2}}(\bar{B}_x + i \bar{B}_y)
		& D - \gamma_e \bar{B}_z
	\end{pmatrix}.
\end{equation}
We define
\begin{align}
	C &= \frac{\gamma_e}{\sqrt{2}} (\bar{B}_x - i \bar{B}_y), \\
	C^\ast &= \frac{\gamma_e}{\sqrt{2}} (\bar{B}_x + i \bar{B}_y), \\
	K &= \gamma_e \bar{B}_z ,
\end{align}
so that
\begin{equation}
	|C|^2 = C C^\ast = \frac{\gamma_e^2}{2}(\bar{B}_x^2 + \bar{B}_y^2).
\end{equation}
The characteristic polynomial is
\begin{equation}
	P(E) = \det(H - E I)
	=
	\begin{vmatrix}
		D + K - E & C & 0 \\
		C^\ast & -E & C \\
		0 & C^\ast & D - K - E
	\end{vmatrix}.
\end{equation}
Evaluating the determinant gives the cubic equation
\begin{equation}
	E^3 - 2 D E^2 + (D^2 - \gamma_e^2 B^2) E
	+ D \gamma_e^2 B_\perp^2 = 0,
\end{equation}
where $B^2 = \bar{B}_x^2 + \bar{B}_y^2 + \bar{B}_z^2$, \qquad $B_\perp^2 = \bar{B}_x^2 + \bar{B}_y^2$.

\subsection{Depressed Cubic Form}

Using the substitution
\begin{equation}
	E = E' + \frac{2D}{3},
\end{equation}
the cubic reduces to the depressed form
\begin{equation}
	E'^3 + p E' + q = 0,
\end{equation}
with
\begin{align}
	p &= -\frac{D^2}{3} - \gamma_e^2 B^2, \\
	q &= \frac{2D^3}{27}
	- \frac{2D \gamma_e^2 B^2}{3}
	+ D \gamma_e^2 B_\perp^2.
\end{align}

\subsection{Eigenvalues}

Since the Hamiltonian is Hermitian, all eigenvalues are real.  
For $p<0$, they can be written as
\begin{equation}
	E_k =
	\frac{2D}{3}
	+ 2 \sqrt{-\frac{p}{3}}
	\cos\!\left( \phi + \frac{2\pi k}{3} \right),
\end{equation}
where
\begin{equation}
	\phi =
	\frac{1}{3}
	\cos^{-1}
	\left(
	\frac{3q}{2p} \sqrt{-\frac{3}{p}}
	\right), \qquad k=0,1,2.
\end{equation}

The ground-state energy $E_g$ is obtained by selecting the minimum of the three real roots in Eq.~(15).  
For $p<0$, the ordering of the cosine terms is fixed by
\[
\cos\!\left(\phi+\frac{4\pi}{3}\right)
\le
\cos\!\left(\phi+\frac{2\pi}{3}\right)
\le
\cos(\phi),
\qquad 0\le \phi \le \frac{\pi}{3},
\]
which implies
\begin{equation}
	\boxed{
		E_g =
		\frac{2D}{3}
		+ 2 \sqrt{-\frac{p}{3}}
		\cos\!\left(\phi+\frac{4\pi}{3}\right), \qquad k=2
	}
\end{equation}
The remaining two roots correspond to the first and second excited states, respectively.




\subsection{Gradient of the Transition Energy}

The sensitivity of a spin transition to magnetic-field fluctuations is quantified by the gradient of the transition energy with respect to the magnetic field. For an arbitrary magnetic field orientation we define the transition energy
\begin{equation}
	f_i = E_{e i} - E_g,
\end{equation}
where $E_g$ and $E_{e i}$ ($i=1,2$) denote the ground and excited eigenenergies, respectively.

The eigenenergies are the real roots of the cubic equation
\begin{equation}
	P(E,\vec{B}) =
	E^3 - 2 D E^2 + (D^2-\gamma^2 B^2)E
	+ D \gamma^2 B_\perp^2 = 0,
	\label{eq:cubic}
\end{equation}
Using implicit differentiation of Eq.~(\ref{eq:cubic}), the gradient of an individual eigenenergy is
\begin{equation}
	\frac{\partial E_k}{\partial b_\alpha}
	=
	-\frac{\partial P/\partial b_\alpha}
	{\partial P/\partial E}
	\bigg|_{E=E_k},
\end{equation}
where
\begin{equation}
	\Lambda_k \equiv
	\frac{\partial P}{\partial E}\bigg|_{E_k}
	=
	3E_k^2 - 4 D E_k + (D^2-\gamma^2 B^2).
\end{equation}
Evaluating the derivatives explicitly yields
\begin{align}
	\frac{\partial E_k}{\partial b_x}
	&=
	-\frac{2\gamma^2 \bar{B}_x (D-E_k)}{\Lambda_k},\\
	\frac{\partial E_k}{\partial b_y}
	&=
	-\frac{2\gamma^2 \bar{B}_y (D-E_k)}{\Lambda_k},\\
	\frac{\partial E_k}{\partial b_z}
	&=
	\frac{2\gamma^2 \bar{B}_z E_k}{\Lambda_k}.
\end{align}
The gradient of the transition energy follows directly as
\begin{equation}
	\vec{\nabla}_{\mathbf{b}} f_i
	=
	\vec{\nabla}_{\mathbf{b}} E_{e i}
	-
	\vec{\nabla}_{\mathbf{b}} E_g .
\end{equation}
To obtain a scalar, orientation-independent measure of first-order magnetic-field sensitivity, we define the gradient magnitude
\begin{equation}
		\mathcal{G}_i
		=
		\left|
		\vec{\nabla}_{\mathbf{b}} f_i
		\right|
		=
		\sqrt{
			\sum_{\alpha=x,y,z}
			\left(
			\frac{\partial f_i}{\partial b_\alpha}
			\right)^2
		}.
	\label{eq:gradient_metric}
\end{equation}
Vanishing $\mathcal{G}_i$ identifies clock or near-clock transitions that are first-order insensitive to magnetic noise.
Evaluating $\frac{\partial f_i}{\partial b_\alpha}$ and substituting them in Eq. \ref{eq:gradient_metric}, leads to

\begin{equation}
		\mathcal{G}_i
		=
		\left|
		\vec{\nabla}_{\mathbf{b}} f_i
		\right|
		= 2 \gamma_e^2
		\sqrt{(\bar{B}_x^2+\bar{B}_y^2)
			\left[
			\frac{(D-E_g)}{\Lambda_g}-\frac{(D-E_{ei})}{\Lambda_{ei}}
			\right]^2
			+\bar{B}_z^2
			\left[
			\frac{E_{ei}}{\Lambda_{ei}}-\frac{E_g}{\Lambda_g}
			\right]^2
		}.
	\label{eq:gradient_metric2}
\end{equation}

\noindent {\bf Parallel field:}

$B_x=B_y=0$, $B_z=B$. The gradient becomes
\begin{equation}
	\left|
	\vec{\nabla}_{\mathbf{b}} f_i
	\right|
	= 2 \gamma_e^2
	\sqrt{(b_x^2+b_y^2)
		\left[
		\frac{(D-E_g)}{\Lambda_g}-\frac{(D-E_{ei})}{\Lambda_{ei}}
		\right]^2
		+\bar{B}_z^2
		\left[
		\frac{E_{ei}}{\Lambda_{ei}}-\frac{E_g}{\Lambda_g}
		\right]^2
	},
		\label{eq:gradient_metric3}
\end{equation}
where $D-E_g=D$ and $D-E_{ei}=\pm\gamma_e B$. Under the approximation $\gamma_e b \ll \gamma_e B\ll D$, $\Lambda_g\approx D^2$ and $\Lambda_{ei}\approx \pm 2D\gamma_e B$. Substituting them in Eq. \ref{eq:gradient_metric3} and simplifying, we get
\begin{equation}
		\left|
	\vec{\nabla}_{\mathbf{b}} f_i
	\right| \approx \gamma_e.
\end{equation}

\noindent {\bf Transverse field:}

$B_y=B_z=0$, $B_x=B$. The gradient becomes
\begin{equation}
	\left|
	\vec{\nabla}_{\mathbf{b}} f_i
	\right|
	= 2 \gamma_e^2
	\sqrt{(\vec{B}_x^2+b_y^2)
		\left[
		\frac{(D-E_g)}{\Lambda_g}-\frac{(D-E_{ei})}{\Lambda_{ei}}
		\right]^2
		+b_z^2
		\left[
		\frac{E_{ei}}{\Lambda_{ei}}-\frac{E_g}{\Lambda_g}
		\right]^2
	},
	\label{eq:gradient_metric4}
\end{equation}
Under the approximation $\gamma_e b \ll \gamma_e B\ll D$, $E_g \approx 0$, $E_{ei}\approx D$, $\Lambda_g\approx D^2$, and $\Lambda_{ei}\approx -\gamma_e^2 B^2$. Substituting them in Eq. \ref{eq:gradient_metric4} and simplifying, we get
\begin{equation}
	\left|
	\vec{\nabla}_{\mathbf{b}} f_i
	\right| \approx 2 \gamma_e^2 B/D.
\end{equation}

\subsection{Curvature of the Transition Energy}

When the first-order magnetic-field sensitivity is suppressed, the dominant contribution to decoherence arises from second-order field fluctuations. This effect is quantified by the curvature of the transition energy with respect to the magnetic field.

Starting again from the implicit eigenvalue condition $P(E_k,\vec{B})=0$, a second differentiation with respect to a field component $b_\alpha$ gives
\begin{equation}
	\frac{\partial^2 E_k}{\partial b_\alpha^2}
	=
	-\frac{1}{\Lambda_k}
	\left[
	\frac{\partial^2 P}{\partial b_\alpha^2}
	+2\frac{\partial^2 P}{\partial E \partial b_\alpha}
	\frac{\partial E_k}{\partial b_\alpha}
	+\frac{\partial^2 P}{\partial E^2}
	\left(\frac{\partial E_k}{\partial b_\alpha}\right)^2
	\right]_{E=E_k}.
	\label{eq:second_derivative_general}
\end{equation}
The required derivatives of $P(E,\vec{B})$ are
\begin{align}
	\frac{\partial^2 P}{\partial b_x^2}
	&= 2\gamma^2 (D-E),\\
	\frac{\partial^2 P}{\partial b_y^2}
	&= 2\gamma^2 (D-E),\\
	\frac{\partial^2 P}{\partial b_z^2}
	&= -2\gamma^2 E,\\
	\frac{\partial^2 P}{\partial E \partial b_\alpha}
	&= -2\gamma^2 b_\alpha,\\
	\frac{\partial^2 P}{\partial E^2}
	&= 6E - 4D.
\end{align}
Substituting these expressions into Eq.~\ref{eq:second_derivative_general} yields closed-form analytical expressions for $\partial^2 E_k/\partial b_\alpha^2$ for all field orientations. The curvature of the transition energy is then
\begin{equation}
	\frac{\partial^2 f_i}{\partial b_\alpha^2}
	=
	\frac{\partial^2 E_{e i}}{\partial b_\alpha^2}
	-
	\frac{\partial^2 E_g}{\partial b_\alpha^2}.
\end{equation}
We define an orientation-independent curvature metric as
\begin{equation}
	\boxed{
		\mathcal{C}_i
		=
		\sqrt{
			\sum_{\alpha=x,y,z}
			\left(
			\frac{\partial^2 f_i}{\partial b_\alpha^2}
			\right)^2
		}.
	}
	\label{eq:curvature_metric}
\end{equation}
The quantity $\mathcal{C}_i$ characterizes the residual second-order magnetic sensitivity of the transition energy, where $\mathcal{G}_i \approx 0$.

\bibliographystyle{apsrev4-2}
\bibliography{bibNV26}